\documentclass[12pt]{article}
\usepackage{epsfig}

\def\Title#1{\begin{center} {\Large {\bf #1} } \end{center}}

\begin{document}

\Title{\bf{Rare Decays} Recent Results from FOCUS (+Theory)}

\bigskip\bigskip


\begin{raggedright}  

{\it Will E. Johns (for the FOCUS collaboration)\index{Johns, W.}\\
Department of Physics and Astronomy\\
Vanderbilt University\\
Nashville, TN USA}
\bigskip\bigskip
\end{raggedright}

\section{Introduction}

The search for rare decays of charm particles is reaching a point where
long range effects should be observable in the next few years. In order
to perform the analysis of such small data sets, several techniques have
become {\it{de facto}} standards in the last decade. In this talk, we
will show new rare charm decay results from the FOCUS experiment. The
analysis has been performed utilizing a bootstrap technique which
we will describe below.

\section{Technique}

The FOCUS analysis technique has emphasized a careful approach to the
treatment of backgrounds in a limited statistics analysis. The
usual approach is to select cuts to optimize signal efficiency relative
to background sidebands. This ``blind'' technique, where the signal
region is masked off, can still lead to a downward 
fluctuation of the sidebands relative to the masked off ``signal'' region
and a more conservative limit on average. 

Further, authors 
frequently use the technique outlined
in reference \cite{Feldman:1997qc} to calculate the confidence levels used in the
calculation of their limits. The approach in \cite{Feldman:1997qc} does not explicitly
include fluctuations in the background. Indeed, the PDG \cite{James:sr}
suggests 
presenting a measure of the experimental sensitivity in addition to the 
reported limit whenever experiments quote a result since none of the 
methods suggested \cite{James:sr}, including \cite{Cousins:1991qz}, properly 
deal with fluctuations in the background and bias in selecting the data.

For our analysis, we chose a method \cite{Rolke:2000ij} which includes 
the background fluctuations directly into the calculation of the likelihood.
The probability of finding a signal rate $\mu$ and a 
background rate $b$ given $x$ events in a signal region and $y$ events
in background sidebands is described by:

$$P_{\mu,b}(x,y)={ {(\mu+b)^xe^{-(\mu+b)}} \over{x!} }~
{ {(\tau b)^xe^{-(\tau b)}}\over{y!} }.$$

Where $\tau$ is the ratio of the number of background events in the
sideband regions to the signal region. The $\tau$ is determined via
Monte Carlo. Rolke and Lopez \cite{Rolke:2000ij} have shown that this method for determining
an upper limit provides better coverage than \cite{Feldman:1997qc}.

Methods presented by Rolke and Lopez have also been shown to reduce 
bias in the selection of optimal cuts \cite{Rolke:2002ix}. Briefly, 
one uses the computed sensitivity on an ensemble of bootstrapped 
(sample with replacement) events from the data. The cuts chosen by 
each selected set in this manner are applied to a second, independent, 
bootstrapped sample.
A sensitivity as well as an upper limit are computed from the second
sample. The sensitivity and branching ratio quoted are the median values 
of the ensemble of results from the second bootstrap.

\begin{figure}[htb]
\begin{center}
\epsfig{file=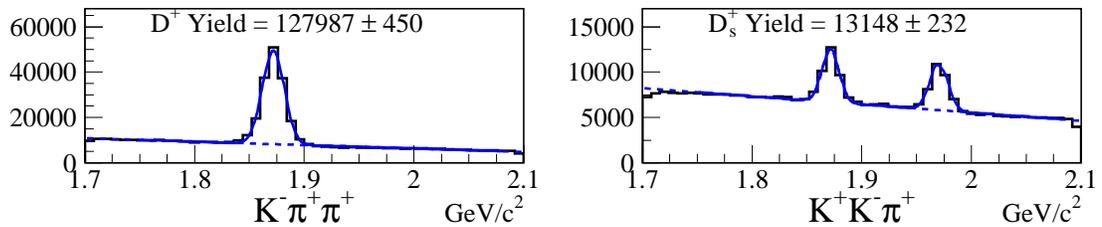,height=1.5in}
\caption{Modes used to normalize the rare decay modes. The yields
are shown for the loosest cut grid combination.}
\label{fig:norm}
\end{center}
\end{figure}

The data for the analyses presented in this talk 
were collected using the Wideband photoproduction experiment FOCUS during the 
1996--1997 fixed--target run at Fermilab. The FOCUS detector is a large aperture, 
fixed--target spectrometer with excellent vertexing and particle identification
used to measure the interactions of high energy photons on a segmented BeO 
target. The FOCUS beamline \cite{Frabetti:1992bn} and 
detector \cite{Frabetti:1990au,Link:2001pg} have been described elsewhere.

To provide a clean sample of
$D^{+}$'s and $D_s^{+}$'s, we look for $D$'s through the
3-body decay chain $D^{+}(D_s^{+})\to h^\mp\mu^\pm\mu^+$ where the
$h$ represents a pion or a kaon. We use a cut grid based on 
kinematic variables and particle ID algorithm results that have been 
shown to be effective for other charm decays. The grid includes 
cuts on the significance of separation between the interaction and decay 
vertices ($L/\sigma$), the confidence level of the vertex fit
of the decay vertex, the confidence level that a particle is 
identified as a muon, the momentum of muon candidates, and
\v Cerenkov likelihoods based on different particle hypotheses
used to separate pions and kaons \cite{Link:2001pg}. The normalizing modes
used to compute the branching ratios for $D^{+}$ and $D_s^{+}$
are shown in Figure \ref{fig:norm} for the loosest cuts in the grid.


\section{Results and Systematic Checks}

The dominant systematic effects in this analysis have been estimated to occur 
from the absolute branching ratios \cite{Groom:in} used to calibrate the normalization 
modes.
The systematic errors, which averaged about 7.6\% for the $D^+$ modes and 
27.6\% for the $D_s^+$ modes, were added to the result using the technique outlined in 
\cite{Cousins:1991qz}.
As a check on the dual bootstrap, another technique was used
that selected a unique cut {\it{set}} based on the results of the cut grid.
The cuts used to determine the best sensitivities in the first bootstrap
are examined for all modes and a best set is determined based on the most
likely cut combination. This cut set is then applied to all modes {\it{once}}
in the spirit of a more traditional ``blind'' analysis. 

The results of the analysis are presented in Table \ref{tab:blood} below. There is 
good agreement between the dual bootstrap, the sensitivity and the single
cut systematic check. Note that the only mode
where the result and the sensitivity show a marked difference is in $D^+ \to K^- \mu^+ \mu^+ $
which might indicate more contamination from $D^+ \to K^- \pi^+ \pi^+ $ than anticipated. 

\begin{table}[htb]
\begin{center}
\begin{tabular}{|l|l|l|l|} \hline 
Decay Mode                    &  Result   & Sensitivity & Single Cut \\ \hline
$D^+ \to K^+ \mu^+ \mu^- $    &  $9.2 \times 10^{-6}$ & $7.5 \times 10^{-6}$ & $11.8 \times 10^{-6}$ \\ \hline
$D^+ \to K^- \mu^+ \mu^+ $    &  $13.1 \times 10^{-6}$& $4.8 \times 10^{-6}$ & $12 \times 10^{-6}$ \\ \hline
$D^+ \to \pi^+ \mu^+ \mu^- $  &   $8.8 \times 10^{-6}$&  $7.6 \times 10^{-6}$&  $7.5 \times 10^{-6}$ \\ \hline
$D^+ \to \pi^- \mu^+ \mu^+ $  &   $4.8 \times 10^{-6}$&  $5.6 \times 10^{-6}$&  $5.2 \times 10^{-6}$ \\ \hline
$D_s^+ \to K^+ \mu^+ \mu^- $  &   $3.6 \times 10^{-5}$&  $3.3 \times 10^{-5}$&  $3.8 \times 10^{-5}$ \\ \hline
$D_s^+ \to K^- \mu^+ \mu^+ $  &   $1.3 \times 10^{-5}$&  $2.1 \times 10^{-5}$&  $2.0 \times 10^{-5}$ \\ \hline
$D_s^+ \to \pi^+ \mu^+ \mu^- $&   $2.6 \times 10^{-5}$&  $3.1 \times 10^{-5}$&  $1.8 \times 10^{-5}$ \\ \hline
$D_s^+ \to \pi^- \mu^+ \mu^+ $&   $2.9 \times 10^{-5}$&  $2.3 \times 10^{-5}$&  $2.2 \times 10^{-5}$ \\ \hline
\end{tabular}
\caption{FOCUS results with incorporated systematic errors for the modes shown. Each number 
represents a 90\% upper limit for the brancing ratio of the decay mode listed.}
\label{tab:blood}
\end{center}
\end{table}

\section{Comparisons to Experiment and Theory}

\begin{figure}[h!]
\begin{center}
\epsfig{file=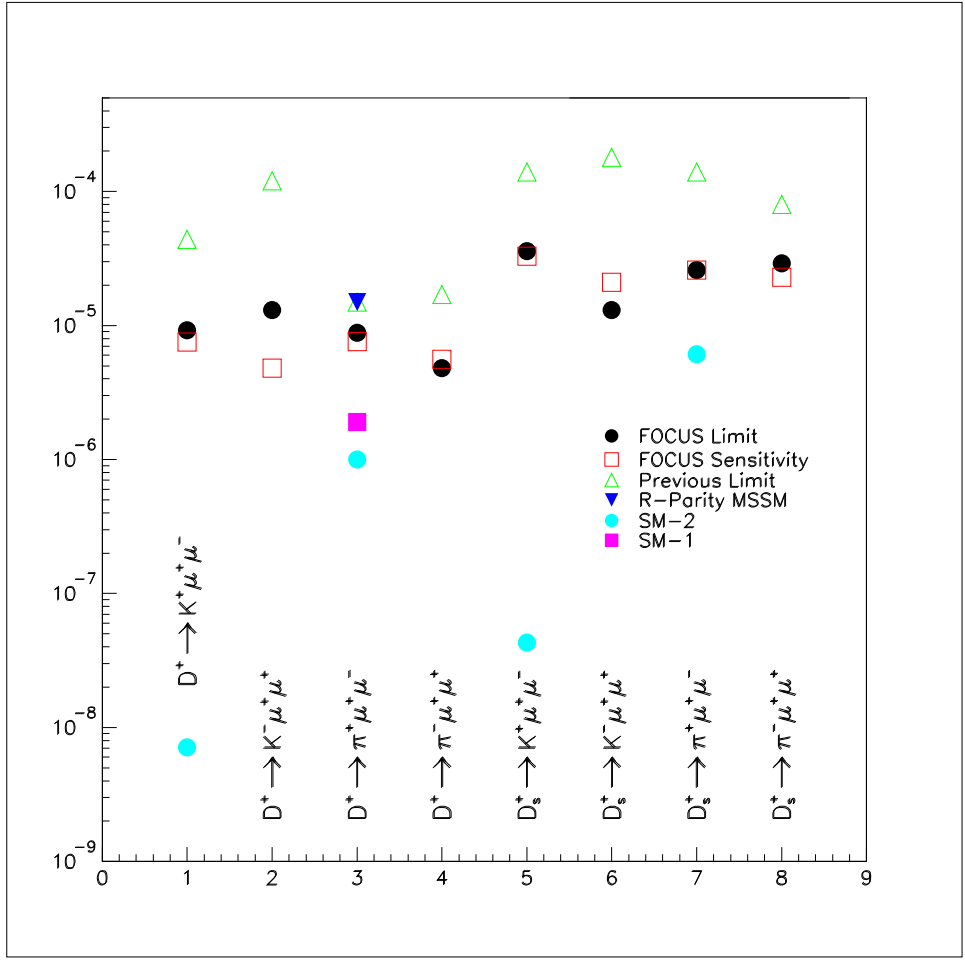,height=5.0in}
\caption{FOCUS results compared to other experiments and theory. The previous  
limits, except for the E687 $D^+ \to K^- \mu^+ \mu^+ $ \cite{Frabetti:1997wp} 
are from Fermilab experiment E791 \cite{Aitala:1999db}. The theory
estimates come from  \cite{Burdman:2001tf} (R-Parity MSSM and SM-1) and \cite{Fajfer:2001sa} (SM-2). 
Note that the SM estimates from \cite{Burdman:2001tf} use a formalism close to
\cite{Singer:1996it}, and at present there is some discrepancy in the invariant
$M_{ll}$ mass behavior for the SM estimates in  \cite{Burdman:2001tf} 
and \cite{Fajfer:2001sa}. The results for
four-body decays measured by E791 \cite{Aitala:2000kk} are not shown.}
\label{fig:comp}
\end{center}
\end{figure}

In Figure \ref{fig:comp} below, we show the FOCUS branching ratio limits for the dual bootstrap
technique described in the text. Our results are a substantial improvement
over previous results \cite{Frabetti:1997wp,Aitala:1999db} and FOCUS sets a new limit for the 
MSSM R-Parity violating prediction \cite{Burdman:2001tf}
for the branching ratio $D^+ \to \pi^+ \mu^+ \mu^- $ of $8.8 \times
10^{-6}~\char'100$  90\% C.L..

\bigskip
I am grateful to Daniel Engh for supplying all the FOCUS results presented in this
note. I am also grateful to Gustavo Burdman and Paul Singer for their patience
during several very useful conversations.

\end{document}